\documentclass{sig-alternate-05-2015}

\usepackage{graphicx} 
\usepackage{subfigure}
\usepackage{algorithmic}
\usepackage{algorithm}
\usepackage{url}
\usepackage{times}
\usepackage{subfigure}
\usepackage{graphicx,epstopdf}
\usepackage{xcolor}
\usepackage{balance}
\usepackage{cite}
\usepackage[english]{babel}
\usepackage{amsmath, bm}

\usepackage{xspace}

\newcommand{\eat}[1]{}
\newcommand{\squishlist}{
 \begin{list}{$\bullet$}
  {  \setlength{\itemsep}{0pt}
     \setlength{\parsep}{3pt}
     \setlength{\topsep}{3pt}
     \setlength{\partopsep}{0pt}
     \setlength{\leftmargin}{2em}
     \setlength{\labelwidth}{1.5em}
     \setlength{\labelsep}{0.5em}
} }
\newcommand{\squishlisttight}{
 \begin{list}{$\bullet$}
  { \setlength{\itemsep}{0pt}
    \setlength{\parsep}{0pt}
    \setlength{\topsep}{0pt}
    \setlength{\partopsep}{0pt}
    \setlength{\leftmargin}{2em}
    \setlength{\labelwidth}{1.5em}
    \setlength{\labelsep}{0.5em}
} }

\newcommand{\squishdesc}{
 \begin{list}{}
  {  \setlength{\itemsep}{0pt}
     \setlength{\parsep}{3pt}
     \setlength{\topsep}{3pt}
     \setlength{\partopsep}{0pt}
     \setlength{\leftmargin}{1em}
     \setlength{\labelwidth}{1.5em}
     \setlength{\labelsep}{0.5em}
} }

\newcommand{\squishend}{
  \end{list}
}

\newcounter{ccc}

\newcommand{\stitle}[1]{\vspace{0.7ex}\noindent{\bf #1}}

\newcommand{\ie}{\emph{i.e.,}\xspace}
\newcommand{\eg}{\emph{e.g.,}\xspace}
\newcommand{\wrt}{\emph{w.r.t.}\xspace}

\newcommand{\kw}[1]{{\ensuremath {\mathsf{#1}}}\xspace}

\newcommand{\ewpr}{{\sc EWPR}\xspace}
\newcommand{\ewprall}{$\kw{EWPR^*}$}
\newcommand{\blpr}{{\sc PR}\xspace}
\newcommand{\blwpr}{{\sc WPR}\xspace}
\newcommand{\blmulrank}{\kw{MulRank}}
\newcommand{\PairAcc}{\kw{PairAcc}}

\begin{document}






%
\CopyrightYear{2016}
\setcopyright{acmcopyright}
\conferenceinfo{WSDM CUP'16,}{February 22, 2016, San Francisco, CA, USA.}
\isbn{XXX-X-XXXX}
\acmPrice{\$XX.XX}
\doi{http://dx.doi.org/XXXX.XXXX}

\title{Ensemble of Time-Weighted PageRank for Literature Ranking}

\title{Entity Ranking with Ensembles in Heterogeneous Graphs}

\title{Entity Ranking with Ensembles}

\title{Ensemble Enabled Time-Weighted PageRank}

\title{Ensemble Enabled Weighted PageRank}

\author{
Dongsheng Luo\hspace{1.5ex} Chen Gong\hspace{1.5ex} Renjun Hu\hspace{1.5ex} Liang Duan\hspace{1.5ex} Shuai Ma$^{*}$\\
\vspace{-.5ex} \\
{\affaddr SKLSDE Lab, Beihang University, China}
\vspace{0ex} \\
{\affaddr \{lds1995, gongchen, hurenjun, duanliang, mashuai\}@buaa.edu.cn }
}

\maketitle
\begin{abstract}
This paper describes our solution for WSDM Cup 2016. Ranking the query independent importance of scholarly articles is a
critical and challenging task, due to the heterogeneity and dynamism of entities involved.
Our approach is called \underline{E}nsemble enabled \underline{W}eighted \underline{P}age\underline{R}ank (\ewpr).
To do this, we first propose Time-Weighted PageRank that extends PageRank by introducing a time decaying factor. We then develop an ensemble method to assemble the authorities of the heterogeneous entities involved in scholarly articles. We finally propose to use external data sources to further improve the ranking accuracy. Our experimental study shows that our \ewpr is a good choice for ranking scholarly articles.
\end{abstract}



\section{Introduction}
\label{sec-intro}


Ranking scholarly articles is a critical and challenging task, due to the heterogeneity and dynamism of entities involved~\cite{fcs-biggraph}.
Generally speaking, a ranking is {\em a function that assigns each entity a numerical score}.
Such a ranking plays a key role in literature recommendation systems, especially in the {\em cold start} scenarios.

This paper focuses on ranking the importance of scholarly articles in a query independent way. As scholarly articles involve with authors, venues, affiliations
and references, they indeed form a complex heterogeneous graph.
Hence, this is essentially a problem of assessing the importance of nodes in a heterogeneous graph.

\begin{figure}
\centering
\includegraphics[scale=0.5]{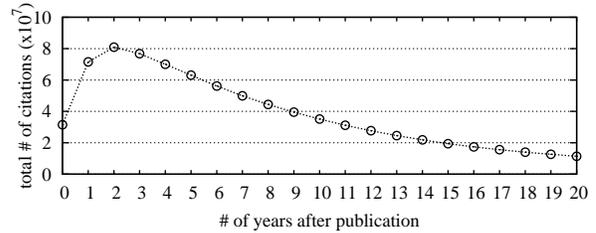}
\vspace{-2ex}
\caption{Citations \wrt the number of published years}
\label{fig-citation}
\vspace{-3ex}
\end{figure}

Comparing with homogeneous rankings such as of Web pages, ranking scholarly articles in a complex heterogeneous graph is much more challenging from two aspects. Firstly, even if we are only to rank one type of entities, \ie scholarly articles, other types of entities such as venues and authors are typically involved. Moreover, the impacts of different types of entities on the ranking of scholarly articles differ from each other. Secondly, entities are evolving, and the importance of an article varies with time~\cite{Li08TSRanking}. Recently published articles are more likely to have increasing impacts in the next few years, and those published many years ago tend to have decreasing impacts since people potentially care more about the latest results. For instance, the citations of articles typically increase in the first two years after publication, and then decrease after that, as shown by the statistics of the Microsoft Academic Graph (MAG)~\cite{Sinha15:MAG} in Figure~\ref{fig-citation}. Indeed, {\em how to accurately rank nodes in heterogeneous graphs remains a challenging task}.


Currently, structure based methods, such as PageRank~\cite{Brin98:PageRank} and its variant Weighted PageRank~\cite{Xing04:WPR}, are among the most effective ones for ranking scholarly articles. However, as pointed out in~\cite{Li08TSRanking}, these previous methods favour older articles that have accumulated a large number of links (\eg citations). However, recently published articles are often underestimated, and they potentially have an increasing impact. These motivate us to develop a new approach to ranking the importance of scholarly articles.


\stitle{Contributions}.
To this end, we propose Ensemble enabled Weighted PageRank (\ewpr) for ranking the importance of scholarly articles, which is among the first to address those challenges above.

\vspace{0.5ex}
\noindent(1) Firstly, we propose Time-Weighted PageRank that extends PageRank \cite{Brin98:PageRank} by introducing a time decaying factor, inspired by Weighted PageRank \cite{Xing04:WPR}, and the authorities of individual articles are discriminately propagated in terms of their own citation information with time, rather than equally propagated like PageRank.

\vspace{0.5ex}
\noindent(2) Secondly, we develop an ensemble method to assemble the authorities of the heterogeneous entities involved in scholarly articles,
which is much more flexible than the mixed model~\cite{Jiang12-MRank} that simultaneously exploits entities and directly produces the ranking.


\vspace{0.5ex}
\noindent(3) Finally, we propose to use external data sources to enrich data and further improve the ranking accuracy.

\stitle{Organization}. The rest of our paper is organized as follows. Section~\ref{sec-model} introduces the ranking model. Section~\ref{sec-impl} discusses how to deal with missing data using external sources. Experimental results are reported in Section~\ref{sec-exp}, followed by conclusions in Section~\ref{sec-conc}.

\section{Ranking Model}
\label{sec-model}

We first introduce our model for ranking scholarly articles.

\subsection{Time-Weighted PageRank}
\label{subsec-twpr}

PageRank~\cite{Brin98:PageRank} has been extensively applied to the ranking of scholarly articles~\cite{Li08TSRanking,Richardson06:BPR,sayyadi09,Zhou07-CoRank}, as hyperlinks among Web pages can be easily replaced with citation relationships among articles, and citation analysis plays a key role to evaluate the importance of scholarly articles. However, the direct use of PageRank for ranking scholarly articles is problematic in terms of the following:

\noindent(1) First, each article equally distributes its authority to its reference articles in the iteration of PageRank~\cite{Brin98:PageRank}, which essentially assumes that each article is equally influenced by its references. However, scholarly articles typically have different impacts in practice, and there is a need to differentiate the impacts of reference articles.


\noindent(2) Second, citation relationships are significantly different from hyperlinks, as the former are time-evolving, and have been successfully exploited in scholarly article ranking~\cite{Li08TSRanking,Wang13AAAI,sayyadi09}. Such temporal information is supplementary to purely structure based PageRank.

\stitle{Time-Weighted PageRank}. We incorporate time information into our ranking model. While in most previous work, time information is simply exploited in the form of exponential decay~\cite{Li08TSRanking,Wang13AAAI,sayyadi09}. We rethink the usage of time information in terms of the impacts of scholarly articles. Recall that Figure~\ref{fig-citation} illustrates the total number of citations \wrt the number of years after the publication of articles.
Here we use the number of citations to evaluate the impacts of articles. As we can see, the number of citations reaches a peak in two years, and gradually decreases after that. This statistical result conforms to our perception of the impacts of articles.


According to Figure~\ref{fig-citation}, the impacts of articles do depend on time, but not simply in the form of exponential decay. Specifically, if an article is cited after the citation peak, its impact should decay with time. Otherwise, its impact is fixed as a constant number, since we argue that the increment of its citations during this period is mainly due to the increase of its popularity.
Moreover, considering that different articles may reach their citation peaks in different ways, we compute the peak time for each individual article, rather than using the same citation peak for all articles.

Inspired by these properties of scholarly article citations, we present Time-Weighted PageRank that evaluates the authorities of nodes in a directed graph, in which each node is attached with time information. It differs from PageRank by weighting the influence propagation using the {\em impact weights on edges}, which represent the relative amounts of authorities that should be propagated from the edge sources to targets, and which also depend on the time information on nodes, following the same temporal tendency as scholarly article citations discussed above.

Formally, the impact weight on directed edge $(u,v)$, \ie edge from $u$ to $v$, is defined as:
\vspace{-1ex}

\begin{small}
\begin{equation} \label{eq-infl-weights}
w(u,v)  =  \begin{cases}  \hspace{10ex} 1 & T_u <  Peak_v \\
  1/(\ln(e+T_u-Peak_v))^t & T_u \geq Peak_v,
\end{cases}
\end{equation}
\end{small}

\vspace{-1ex}
\noindent
where $T_u$ is the time information on node $u$, $Peak_v$ is the peak time of node $v$ using the time information of all nodes connecting to $v$, and $t$ is the decaying factor. By default,  Eq.~(\ref{eq-infl-weights}) uses years as its time granularity. For the sake of completeness, we further set $w(u,v)$ to $0$ if these does not exist an edge from $u$ to $v$.

The authority update rule in Time-Weighted PageRank is:
\vspace{-1ex}

\begin{small}
\begin{equation}\label{eq-twpr}
PR(v)=(1-d)+d \cdot \sum_{u\in IN(v)} \frac{w(u,v)\cdot PR(u)}{W(u)},
\end{equation}
\end{small}

\vspace{-1ex}
\noindent where $PR(u)$ and $PR(v)$ are the authorities of $u$ and $v$, respectively, $IN(v)$ is the set of nodes having edges to $v$, $W(u)=\Sigma_{v} w(u,v)$ is the sum of impact weights on all edges from $u$, and $d$ is a damping parameter in $[0, 1]$. From Eq.~(\ref{eq-twpr}) we can see that authorities are based on the impact weights, not equally distributed.

\stitle{Remarks}. Note that here Eq.~(\ref{eq-twpr}) is indeed a more general update rule than Weighted PageRank~\cite{Xing04:WPR}, and the name of Time-Weighted PageRank comes from the use of time information in the initial impact weight  $w(u,v)$ of Eq.~(\ref{eq-infl-weights}).

\subsection{Ensembles}
\label{subsec-ensemble}

We start this part by thinking about how people evaluate the importance of scholarly articles. In practice, the importance of an article can be evaluated according to many factors such as citations, venues and authors. Only focusing on the citation information limits the accuracy of the results. Consider the case when we are to evaluate a newly published article whose citations are not currently available. In this case citation information fails to give a reasonable rank, but other information such as venues and authors could be used instead to refine the rank. Hence, we propose the use of an ensemble model, in which each ensemble is essentially a ranking based on the authorities of one type of heterogeneous entities, and these ensembles are assembled to produce the final ranking.

\stitle{Citation ensemble}.
The first ensemble is based on the authorities of articles and it is called citation ensemble since we use citation information to evaluate these authorities.
Specifically, it first uses citation information to construct a directed graph, where a node represents an article and an edge $(u,v)$ denotes that $u$ cites $v$ as its reference. The graph is further associated with time information such that (1) the publication years of articles are attached to corresponding nodes, and (2) the peak time of each node is the year with the largest number of citations, in which ties are broken randomly if existing.
After that, Time-Weighted PageRank is run on the graph and each node is assigned its authority. Finally, the ensemble maps each article to the authority of its corresponding node as its rank.


\stitle{Venue ensemble}.
The second ensemble is based on the authorities of venues. It first evaluates the authority of each venue, and then maps each article to the authority of the venue where it is published as its rank.
To do this, we also construct a directed graph, in which a node represents a venue and an edge $(s,t)$ means that there is at least one article published in $s$ citing at least one article published in $t$. We also use {\em impact weights} to denote the weights among venues.
And the impact weights are defined as sums of impact weights between articles published in the corresponding venues:
\vspace{-1ex}

\begin{small}
\begin{equation} \label{eq-infl-weights-v}
w_v(s,t)  = \sum_{u\in C(s), v\in C(t)} w(u,v).
\end{equation}
\end{small}

\vspace{-1ex}
Here, $C(s)$ and $C(t)$ are the collections of articles published in venues $s$ and $t$, respectively, and $w(u,v)$ is the impact weight of articles $u$ and $v$ produced in the citation ensemble. It then iteratively computes the authorities of venues using the impact weights of venues and the update rule in Eq.~(\ref{eq-twpr}).

\stitle{Author ensemble}.
The third ensemble is based on the authorities of authors. Similar to the venue ensemble, we could first evaluate the authority of each author and then map each article to the average authority of the author(s) associated with the article as its rank.
However, the resulting graph is too large to handle. Hence, we adopt another way to evaluate the authority of an author, by using the average authority of all articles published by the author, which are produced by the citation ensemble.

\stitle{Affiliation ensemble}.
Recall that articles in our data are also associated with affiliation information. Following the way of the venue or author ensemble, we can derive another ensemble, \ie affiliation ensemble. However, we argue that the use of affiliation ensemble may have negative effects since the correlation between the importance of an article and the average authority of its affiliation(s) is not as strong as others such like authors and venues. As shown by the experimental study in Section~\ref{sec-exp},  the incorporation of the affiliation ensemble impairs the ranking accuracy. Hence, we choose not to use the affiliation ensemble in our model.

\stitle{Remarks}.
Traditional PageRank equally distributes the authorities of nodes, and PageRank based models suffer from the problem that older articles are preferred since they have accumulated a large number of citations~\cite{Li08TSRanking}, and Time-Weighted PageRank based models alleviate the problem to a certain degree by lowering the impact weights of articles when they are cited after their peak time, \ie $T_u\geq Peak_v$. We further propose the venue and author ensembles to improve the ranking accuracy.



\subsection{Ensemble Enabled Ranking}
\label{subsec-eerank}

The aforementioned ensembles (except the affiliation one) are finally assembled to produce the final ranking, referred to as \underline{E}nsemble enabled \underline{W}eighted \underline{P}age\underline{R}ank (\ewpr). Before assembling, each ranking is properly scaled such that the average scores of different rankings are the same. Suppose that the scaled ranking scores of articles $u$ are $R_c(u)$, $R_v(u)$, and $R_a(u)$ from the citation ensemble, venue ensemble and author ensemble, respectively. The final ranking score of $u$ is aggregated as follows:
\vspace{-1ex}

\begin{small}
\begin{equation} \label{eq-ensemble}
R(u) =  \frac{R_c(u) + \alpha \cdot R_v(u) + \beta \cdot R_a(u)} {1+\alpha+\beta}.
\end{equation}
\end{small}

\vspace{-1ex}
\noindent Here parameters $\alpha$ and $\beta$ as well as the value $1$  are used to regularize the contributions of the citation, venue and author information. Intuitively, these values indicate the intensity of the correlation between the importance of articles and the specific information.


\begin{figure}
\centering
\includegraphics[scale=0.12]{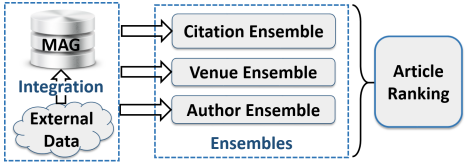}
\vspace{-2ex}
\caption{Architecture of our ranking model \ewpr} \label{fig-framework}
\vspace{-3ex}
\end{figure}

We close this section by presenting the architecture of our ranking model \ewpr, illustrated in Figure~\ref{fig-framework}.
Our model \ewpr contains three distinct ensembles, \ie the citation ensemble, venue ensemble and author ensemble. The citation ensemble directly uses Time-Weighted PageRank, while the other two are partially based on Time-Weighted PageRank. These ensembles are further assembled to produce the final ranking. As illustrated in Figure~\ref{fig-framework}, external data is also exploited in \ewpr. How to collect and use external data will be introduced in the coming section.

\section{Dealing with Missing Data}
\label{sec-impl}

Data quality is one of the most challenging issues in large scale data management, especially for data from open domains and multiple sources, \eg the Microsoft Academic Graph (MAG)~\cite{Sinha15:MAG}.
The early version of MAG has $120$ million scholarly articles, among which we find that there are about $73$ million articles without references and about $77$ million ones without venues. The ranks of those articles with missing information are underestimated by our model \ewpr, since ensembles assign the minimum scores to articles. As a result, data missing seriously impairs the ranking accuracy.

As for references and venues, the later are easier to obtain, and each filled venue can have a direct and substantial impact on the article ranking, \ie $R_v(u)$ of Eq.~(\ref{eq-ensemble}). In contrast, a filled reference only has an indirect and slight impact. Hence, we decide to use external data to fill in missing venues.

\stitle{Data collecting}.
The raw external data is collected from publicly available Digital Libraries, such as IEEE Xplore ({\footnotesize http://ieeexplo-re.ieee.org/gateway/}),  PubMed ({\footnotesize http://www.ncbi.nlm.nih.gov/pub-med/}) and DBLP ({\footnotesize http://dblp.uni-trier.de/db/}). In total, we collect $2.8$ million articles with venue information as our external data, in which there are $57,000$ different venues.

\stitle{Data preprocessing}.
The venues in MAG are well processed, and are replaced by their series names. For example, {\em ``9th International Conference on Web Search and Data Mining, 2016''} is replaced with {\em ``Web Search and Data Mining''}. This makes it hard to directly link with the collected raw venue names. Hence, we preprocess raw venue names for the simplification of subsequent venue linking.
We first remove stop words such as {``on''} and common words like {``Conference''}, as well as years and some special characters from collected raw venue names. Then the same venues are merged, and the number of different venue names is reduced to $42,000$.

\stitle{Data linking}.
The final and also the most important step of filling missing venue information is to link each collected venue name to an existing one in MAG. Intuitively, linking based on name similarity is the most effective way such that two venues are linked if their names bear high similarity. We exploit the Jaro metric to evaluate the name similarity, which is based on the number and order of the common characters between two strings, and obtains good results in tasks such as record linkage and name matching~\cite{Cohen03strcompa}. Formally, a collected venue name is linked to an existing one in MAG if their Jaro similarity exceeds a pre-define threshold.

However, such a threshold is nontrivial to determine in practice. A high threshold can guarantee the accuracy of linked pairs, while only a tiny proportion of collected venue names are linked. On the other hand, a low threshold increases the number of linked pairs, which, in the same time, also introduces many errors. In order to reach a good balance between the number of linked pairs and the accuracy, we propose to combine another constraint on topic similarity of venues for linking, and only weaker filter conditions need be used in both constraints.

In MAG, fields of study (FOS) represent research topics of articles, such as {\em Web pages}, and {\em language technology}. Hence, we use FOS to evaluate the topic similarity of two venues. There are about $54,000$ FOS in MAG and most articles are assigned with two or three FOS. Let the set of FOS of each venue be the union of the sets of FOS of articles published in that venue. And the topic similarity of two venues based on FOS is defined as:
\vspace{-1ex}

\begin{small}
\begin{equation} \label{eq-fos}
TS(s,t)=({|F_s\bigcap F_t|})/{\sqrt{|F_s|\cdot|F_t|}},
\end{equation}
\end{small}

\vspace{-1ex}
\noindent in which $s$ and $t$ are two venues, and $F_s$ and $F_t$ are the sets of FOS of $s$ and $t$, respectively.

When we link a collected venue name, it is directly linked to the most similar one in terms of name similarity, if their Jaro similarity exceeds a high threshold $\lambda$. Otherwise, we first use the topic similarity constraint to select several candidates in MAG, \ie venues whose topic similarities with the collected venue exceed a threshold $\theta$. Intuitively, these candidates are in the similar fields of the collected one. We then select the most similar candidate in terms of name similarity as its linked venue, if their Jaro similarity exceeds another threshold $\phi$. Hence, the collected venue is linked to the one to which it is similar in terms of both topics and names.

In our model \ewpr, threshold $\lambda$ is set to $0.95$, while thresholds $\theta$ and $\phi$ need not be very high, which are $0.5$ and $0.7$, respectively. Finally, $6,000$ among the $42,000$ collected venues are linked, resulting in $340,000$ (about $12\%$) articles with enriched venue information. Note that a majority of the collected venue names are not valid venues, such as booktitles and names of workshops, and cannot be linked to any one in MAG.

\section{Experimental Study}
\label{sec-exp}

In this section, we use the Microsoft Academic Graph~\cite{Sinha15:MAG} to evaluate the effectiveness of our ranking model \ewpr in terms of four aspects: (1) Time-Weighted PageRank {\em vs.} PageRank, (2) single ensembles {\em vs.} multiple ensembles, (3) ensemble models {\em vs.} mixed models and (4) ensemble models with affiliations.

\subsection{Experimental Settings}

We first present our settings.

\eat{
\begin{table}[t!]
\label{tab-statistics}
\begin{center}
\begin{scriptsize}
\vspace{1ex}
\begin{tabular}{|c|c|c|}
\hline
{\bf Entity / Relation}       &  {\bf Quantity in Phase~1}     & {\bf Quantity in Phase~2} \\
\hline\hline
Paper      &  $122,675,085$       &  $120,887,833$ \\ \hline
Author      &  $123,017,488$       &  $119,892,201$ \\ \hline
Venue      &  $24,841$       &  $24,843$ \\ \hline
Affiliation      &  $2,716,493$       &  $19,849$ \\ \hline
Fields of study     &  $53,834$       &  $53,830$ \\ \hline
Reference      &  $757,462,733$       &  $952,364,264$ \\ \hline
P-A      &  $324,948,062$       &  $312,034,259$ \\ \hline
P-V      &  $45,783,880$       &  $45,290,168$ \\ \hline
\end{tabular}
\vspace{-5ex}
\end{scriptsize}
\end{center}
\caption{Statistics of MAG}
\vspace{-3ex}
\end{table}
}

\stitle{(1) Dataset}.
The Microsoft Academic Graph (MAG) dataset is a heterogeneous graph containing different types of literature entities and relationships. Please refer to~\cite{Sinha15:MAG} for more details about MAG.

\stitle{(2) Metric}.
We adopt {\em pairwise accuracy}~\cite{Richardson06:BPR} to evaluate the ranking quality, which is the fraction of times that a ranking agrees with the correct importance orders of scholarly article pairs:
\vspace{-1ex}

\begin{small}
\begin{equation}
\label{eq-metric}
\PairAcc=\frac{\#\mbox{ of agreed pairs}}{\# \mbox{ of all pairs}}.
\end{equation}
\end{small}

\vspace{-1ex}
The ground truth is generated by human experts, who are asked to give the orders of importance of article pairs ({\footnotesize https://wsdmcupchall-enge.azurewebsites.net/Home/Rules}).

\stitle{(3) Baselines}.
We compare our method \ewpr, with four baseline methods: \blpr, \blwpr, \ewprall, a variant of \ewpr that further uses the affiliation ensemble, and \blmulrank~\cite{Jiang12-MRank}.

\blpr simply runs PageRank on the citation network, and uses the authority scores as the importance of articles,
\blwpr is only based on the results of the citation ensemble, 
\blmulrank uses the mixed model, where entities are exploited simultaneously, and
\ewprall\ further uses the affiliation ensemble on the basis of \ewpr.

\stitle{(4) Implementation}. For all algorithms: (1) the number of iterations is set to $30$, and (2) the damping parameter $d$ and decaying factor $t$ are fixed to $0.15$ and $2.5$, respectively.
\blmulrank uses the default parameters recommended in~\cite{Jiang12-MRank}.
We further fix $\alpha=1.2$ and $\beta=0.3$ for \ewpr and \ewprall.
And the weight of affiliation ensemble for \ewprall is also set to $0.3$.

All experiments were run on a PC with 2 Intel Xeon E5--2630 2.4GHz CPUs and 64 GB of memory.

\subsection{Experimental Results}
\label{subsec-expres}

We next present the experimental results, which were mainly tested using the training data of WSDM Cup Phase~1. And all experiments were tested without using the external data, which is for Phase~2. The results are reported in Table~1.

\stitle{(1) Time-Weighted PageRank {\em vs.} PageRank}. We first compare the effectiveness of Time-Weighted PageRank with PageRank. The pairwise accuracy of \blpr and \blwpr is $0.687$ and $0.701$, respectively. And our model \blwpr outperforms \blpr by $1.4\%$, which is achieved by introducing a time decaying factor. The results show that Time-Weighted PageRank combining both time and structural information is a better choice for ranking scholarly articles.

\stitle{(2) Single ensembles {\em vs.} multiple ensembles}. We then compare the effectiveness of using single ensemble, \ie the citation ensemble, with multiple diverse ensembles. The pairwise accuracy of \blwpr and \ewpr is $0.701$ and $0.733$, respectively. \ewpr outperforms \blwpr by $3.2\%$ since it combines the citation, venue and author ensembles to evaluate importance of articles. Hence, multiple ensembles are typically more effective than single ensembles.

\stitle{(3) Ensemble models {\em vs.} mixed models}. We also compare the effectiveness of the ensemble model with the mixed model. The pairwise accuracy of \blmulrank and \ewpr is $0.699$ and $0.733$, respectively.
And \ewpr is better than \blmulrank by $3.4\%$.  Moreover, \blmulrank is even worse than the single ensemble method \blwpr. We believe that the effectiveness of the mixed model is impaired by error propagation in a noisy dataset, whereas
the ensemble model controls the impacts of error propagation to some extent. Combining with the previous one, we claim that in noisy heterogeneous graphs, models assembling multiple ensembles are the best choice.

\stitle{(4) Ensemble models with affiliations}. In the last set of tests, we compare \ewpr with \ewprall to evaluate the effectiveness of ensemble models with affiliations. The pairwise accuracy of \ewpr and \ewprall is $0.733$ and $0.711$, respectively. And the incorporation of the affiliation ensemble decreases the pairwise accuracy by $2.2\%$. The results verify our early claim that the correlation between the importance of articles and authorities of affiliations is not as strong as the other three. Moreover, different types of entities may have different contributions to the ranking of articles. Simply combining all information may not be the best way.


Finally, the accuracy of \ewpr in the Leaderboard of Phase~1 is $0.656$ ({\footnotesize https://wsdmcupchallenge.azurewebsites.net/Home}).

\eat{
\stitle{Summary}.
From these tests we find the following.

\noindent(1) The Time-Weighted PageRank which weights the propagation of authority achieves good effectiveness for ranking scholarly articles, \ie outperforming the traditional PageRank by $1.4\%$ in terms of pairwise accuracy.

\noindent(2) The ensemble model is better than using citation information alone and the mixed model by $3.2\%$ and $3.4\%$ in pairwise accuracy, respectively. We believe that the ensemble model is the best choice in the scenario of noisy heterogeneous graphs.

\noindent(3) Different types of literature entities may have different contributions to the ranking of scholarly articles. Simply combining all information does not mean the best result, \eg affiliation information damages the effectiveness.
}

\begin{table}[t!]
\label{tab-result}
\begin{center}
\begin{scriptsize}
\vspace{1ex}
\begin{tabular}{|c|c|c|c|c|c|}
\hline
{\bf Methods}   &  \blpr     & \blwpr  &  \blmulrank  &   \ewpr  & \ewprall    \\
\hline
${\PairAcc}$  & $0.687$   & $0.701$   & $0.699$     & {\bf 0.733}     & $0.711$    \\ \hline
\end{tabular}
\vspace{-4.5ex}
\end{scriptsize}
\end{center}
\caption{Results of pairwise accuracy}
\vspace{-3ex}
\end{table}

\section{Conclusions}
\label{sec-conc}
We have proposed a novel model for scholarly article ranking, which combines Time-Weighted PageRank and ensembles. The authority propagation of scholarly articles are weighted based on each article's individual citation information with time, and the ranking of articles is given by assembling the results of citation, venue and author ensembles. We have also proposed to use external data to enhance the quality of  data for improving the ranking accuracy.



\balance
\stitle{Acknowledgments}.
This work is supported in part by  973 program ({\small No. 2014CB340300}), NSFC ({\small No. 61322207\&61421003}), and MSRA Collaborative Research Program. We also thank team members Dr. Xuelian Lin and Niannian Wu for their support.
For any correspondence, please refer to Shuai Ma.

%

\vspace{1ex}
\bibliographystyle{abbrv}
\begin{small}
\bibliography{sigproc}  
\end{small}
%
%

\end{document}